# Large-scale fission data generation with BSkG3

*Adrián* Sánchez-Fernández [1*], *Wouter* Ryssens[1], and *Stéphane* Goriely[1]

[1]Institut d'Astronomie et d'Astrophysique, Université Libre de Bruxelles. Brussels, Belgium.

**Abstract.** Modeling fission properties, such as barriers and rates, is highly challenging. The most microscopic methods available are based on energy density functionals (EDFs) and rely on a limited set of collective coordinates to describe the evolution of a fissioning nucleus from its ground state to scission. Leveraging the efficiency of the MOCCa nuclear structure code and the predictive power of the BSkG3 EDF, we systematically study fission properties of the heaviest nuclei (roughly 3,300) accounting for (1) axial, triaxial and octupole moment; (2) all nuclei, including odd and odd-odd systems; and (3) fission paths determined via the least-action principle. We present the set of primary fission barriers and spontaneous fission half-lives we obtain and discuss their implications for r-process nucleosynthesis.

## 1 Introduction

Since its discovery, fission has always been an active field of research both regarding its purely theoretical challenge and its practical applications. Almost all existing evaluations of fission probabilities rely on phenomenological models whose free parameters were adjusted to reproduce experimental data. Although such adjustments respond to the needs of some applications, their predictive power remains poor due to the large numbers of free parameters. Such methods should be avoided in applications requiring a purely theoretical description of fission for experimentally unknown nuclei. This is the case of nuclear astrophysics, where reliable predictions are required for heavy, exotic nuclei near the neutron drip line. In this context, one must determine fission barriers (or more generally, fission paths) along with nuclear level densities at the saddle points and fission fragment distributions.

To date, only a limited number of nuclear models have been applied to large-scale predictions of fission rates in the region relevant to the rapid neutron-capture process (or r-process) [1], namely for nuclei heavier than Th [2-5]. Most existing studies rely on semi-empirical microscopic-macroscopic (mic-mac) models, which generally fail to reproduce simultaneously ground state (GS) and fission properties. In contrast, microscopic approaches based on the Hartree-Fock-Bogoliubov (HFB) theory can offer a unified description of both. Among these, the BSkG3 parametrization stands out, achieving an rms deviation of 0.63 MeV for 2457 nuclear masses and rms deviations of 0.33 (0.51) MeV for the 45 empirical primary (secondary) fission barriers [6]. Recent work has further shown that BSkG3 reproduces experimental spontaneous fission (SF) half-lives within three orders of magnitude, an accuracy comparable to that of the most successful highly parameterized mic-mac models [7].

Motivated by the good agreement with experiment, we performed a large-scale survey of fission properties in the region $Z = [80,118]$ encompassing all isotopes from the proton to the neutron-drip line predicted by BSkG3. This is the first large-scale campaign to explore explicitly axial, triaxial and octupole deformed configurations for both even and odd-mass nuclei within a microscopic framework. We obtained a set of fission paths, microscopic collective masses and level densities at the saddle points that are consistent with the description of GS properties. Furthermore, we obtained the SF half-lives for all the nuclei studied, a key quantity that limits the production of super-heavy elements during the r-process nucleosynthesis [5]. All these properites were included in the latest version of TALYS [8] and work on the rest of relevant nuclear inputs for the r-process (n-induced and $\beta$-delayed fission rates and fission fragment distributions) is currently in progress.

## 2 Methodology

While fission is an intrinsically time-dependent process, under the assumption of adiabaticity, one can decouple the motion of the single-particle degrees of freedom from the collective motion of the fissioning nucleus. In the context of fission modelling with EDFs, this implies that fission properties can be derived from the so-called potential energy surface (PES), which maps the energy of the nucleus to a set of collective coordinates that describe how the nucleus evolves from its GS to scission. Among these collective coordinates, the common practice is to choose a few mass multipole moments

$$Q_{lm} = \int d^3r \rho(\boldsymbol{r}) r^l Y_{lm}(\theta,\phi) \qquad (1)$$

---

* Corresponding author: adrian.sanchez.fernandez@ulb.be

that describe the shapes that appear during the fission process. These multipole moments depend on the nuclear matter density $\rho(\mathbf{r})$ and the usual spherical harmonics, $Y_{lm}$. In this regard, the axial quadrupole moment ($Q_{20}$) accounts for the elongation of the nucleus, while the axial octupole moment ($Q_{30}$) allows us to describe asymmetric fission. In addition, it is well-known that allowing for triaxial shapes can lower the inner fission barrier in the actinide region [9].

Even though higher-rank multipole moments or pairing correlations also evolve during fission, exploring additional collective coordinates increases the dimensionality of the PES, thereby raising substantially the computational cost. Although such complex multi-dimensional analyses have been done recently [10,11], they were applied only to a few case studies, and its systematic application for thousands of nuclei remain prohibitive.

In this work we used the MOCCa nuclear structure code [12] to solve the constrained HFB equations using the BSkG3 model. MOCCa solves the HFB equations in a three-dimensional coordinate-space representation, which preserves the numerical accuracy independently of the shape of the nucleus, provided the simulations box is big enough to accommodate the nucleus.

We applied the two-step methodology previously used for our SF half-lives survey [7]. In a first step, we compute an initial PES in $(Q_{20}, Q_{30})$ on which we obtain the minimum energy path (MEP). This path allows us to obtain the energy-optimum octupole shape for each quadrupole moment. In a second step, we compute a PES in $(Q_{20}, Q_{22})$, imposing an additional constraint in $Q_{30}$ that depends on $Q_{20}$ as given by the previous MEP. As a final step, we find the path that minimizes the action integral

$$S = \frac{1}{\hbar} \int_{s_{\rm in}}^{s_{\rm out}} ds \sqrt{2\mu(s)[V_{\rm eff}(s) - E_{\rm in}]} \quad (2)$$

in which the points $s_{\rm in}$, $s_{\rm out}$ are the semiclassical inner and outer turning points of the tunnelling process ($E = E_{\rm in}$); $\mu$ is the collective inertial mass derived from the microscopic inertia tensor in the adiabatic-time-dependent HFB approximation; $V_{\rm eff}$ is the HFB energy and $E_{\rm in}$ is the GS energy of the compound nucleus. Both MEP and least-action path (LAP) are obtained using the PyNEB code [13]. The action associated to the LAP is then used to compute the SF half-lives in the WKB approximation as

$$t_{1/2}^{\rm SF} = \frac{\ln 2}{n}[1 + e^{2S}] \quad (3)$$

Where $n = 10^{20.38}$ s$^{-1}$. Additional details can be found in [7] and the numerical aspects of our calculations will be published elsewhere [14].

In Fig. 1 we present the results for $^{240}$Pu as a showcase. We used the dimensionless deformation parameters $\beta_{lm} = \frac{4\pi}{3R^l} Q_{lm}$, with $R = 1.2A^{1/3}$ fm. In the upper plot we show the PES in $(\beta_{20}, \beta_{22})$ along with the LAP from the GS to the semiclassical outer turning point. In the middle panel we show the energy (black line) and the microscopic collective inertia (red line). We observe an excellent agreement with the empirical fission barriers and isomer RIPL-3 reference values [15], with an accuracy with an accuracy of about 300 keV for the three points. Notice that the inertia is generally decreasing with increasing $\beta_{20}$ and exhibits peaks due to the abrupt changes in triaxiality (near $\beta_{20} = 0.25$ and $0.60$). Thanks to the two-step methodology, we can include the energy-optimal octupole deformation at each $\beta_{20}$ (lower panel) which is in good agreement with the well-known fact that only the outer fission barrier presents a left-right asymmetry for this nucleus. Moreover, the SF half-life obtained with BSkG3 is $2.4 \cdot 10^{19}$ s, while its experimental value is $3.6 \cdot 10^{18}$ s [16].

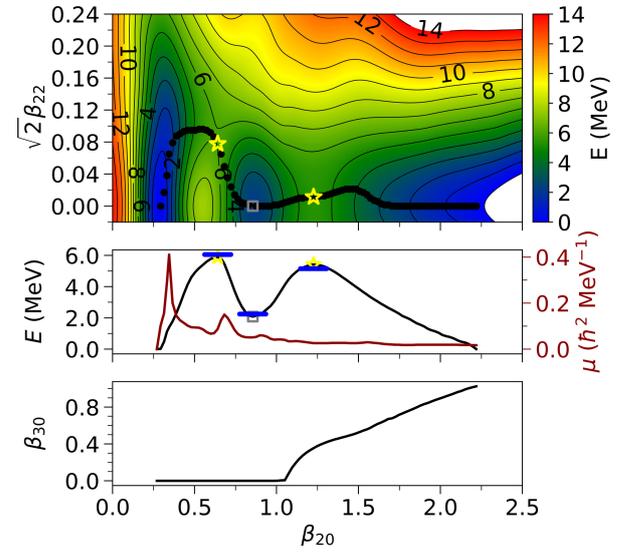

**Fig. 1.** Upper panel: $^{240}$Pu PES with the LAP (black dots) and the saddles/wells (yellow stars/grey squares). Middle panel: Energy (black line) and microscopic collective inertia (red line) along the LAP; Blue lines denote the empirical values from RIPL-3 [15]. Lower panel: evolution of the axial octupole deformation along the LAP.

This level of agreement for barriers is superior to that achieved by other microscopic models previously employed in r-process simulations. As illustrated in Fig. 2. we compare the predictions of HFB-14 [17], BCPM [18] and BSkG3 w.r.t to the 45 empirical values compiled in RIPL-3 [15]. The rms deviations found for these three models are 0.601, 1.419 and 0.32 MeV for HFB-14, BCPM and BSkG3, respectively. In contrast to the other models, BSkG3 incorporates these empirical barriers in its fitting protocol, and the fission paths are determined by exploring both triaxial and octupole degrees of freedom.

This scheme differs from BCPM, which is limited to axial quadrupole deformations, and from HFB-14, which incorporates triaxial effects only through phenomenological corrections on top of the axial fission barriers.

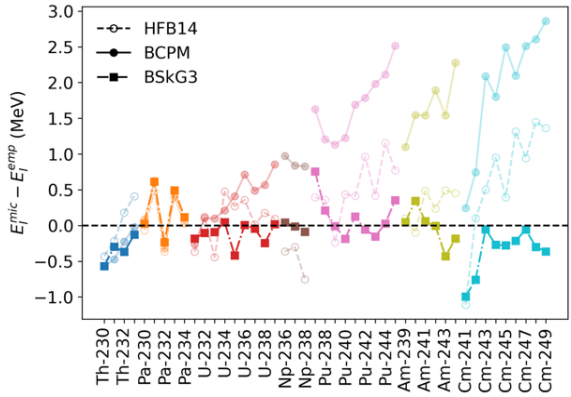

**Fig. 2**. Deviation of the primary fission barriers, $E_I$, obtained with HFB-14, BCPM and BSkG3 microscopic models wr.t. the RIPL-3 empirical values. Colors denote different isotopic chains.

In addition to the agreement with empirical fission barriers and SF half-lives, BSkG3 predictions also fairly well reproduce experimental fission fragment distributions. To ensure consistency with the BSkG3 model, we perform time-dependent generator coordinate method calculations based on our $(Q_{20}, Q_{30})$ surfaces using the FELIX code [19]. A fully quantitative comparison with experimental data requires the inclusion of prompt neutron emission, which we do not currently account for; this will be addressed in future work. Nevertheless, as shown in Fig. 3, the pre-neutron emission yields already reproduce the peak structure and provide a reasonable description of the widths observed experimentally [20,21]. To facilitate comparison, we apply a double Gaussian folding in Z and A with widths $\sigma_Z \approx 2.0$ and $\sigma_A \approx 5.5$, consistent with the typical neck size of our scission configurations (4-6 particles/fm³) and the mass of the nuclei.

## 3 Large-scale deployment

Thanks to the numerical efficiency and robustness of MOCCa, we were able to compute around 6,600 PES covering the region $Z \in [80,118]$. In Fig. 4 (a) we show the primary fission barriers, $B_f$, while panels (b) and (c) show their difference with the BSkG3 neutron separation energy and the $\beta^-$ $Q$-value respectively. This allows us to assess the stability with respect to n-induced and $\beta$-delayed fission channels.

Fig. 4 (b) shows that nuclei around $^{254}$Bk ($Z = 97$) are susceptible to experience n-induced fission ($B_f - S_n \approx 0$), which may hinder the production of $^{254}$Cf via $\beta$-decay and hence affect the hypothesized contribution to the late-time decay heat in neutron star merger ejecta [22].

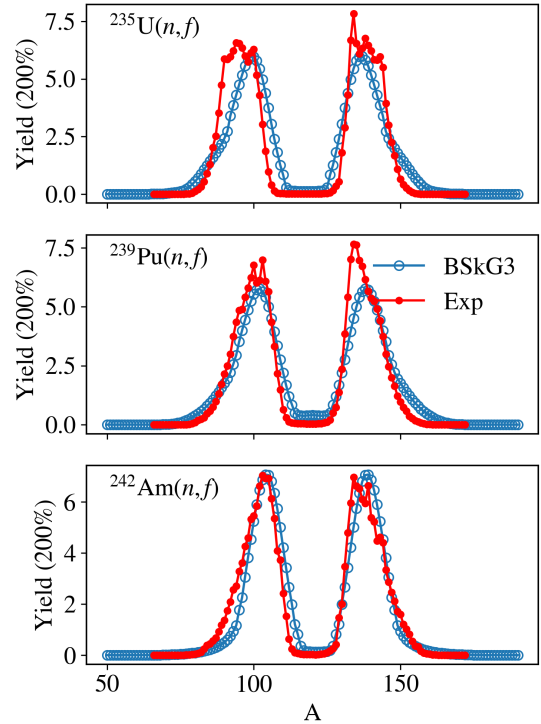

**Fig. 3**. Fission fragments distributions for selected n-induced fission reactions in the actinide region obtained with BSkG3 (blue) and from experiment (red).

Fig. 4 (c) indicates that nuclei near the n-drip line are highly prone to experience $\beta$-delayed fission ($B_f - Q_\beta < 0$). However, the precise termination point of heavy-element synthesis requires a detailed comparison of the relevant fission rates and detailed r-process calculations; this is left for future work.

Additionally, Fig. 5 presents the SF half-lives obtained with (a) BSkG3 and (b) the empirical formulas from [23], that rely on the primary fission barriers obtained with ETSFI [24], with FRLDM [3] values adopted only when ETFSI data is unavailable. Given the wide range of values, we display their decimal logarithm instead. From Fig. 5 (a) we see that near $Z = 108, N = 230$ the SF half-lives reach values the order of $10^{-2}$ s, thereby directly competing with the β-decay timescales. This makes this region a potential termination point for the synthesis of heavy elements and a likely source of fission fragments contributing to the fission-recycling process.

When compared with the empirical systematics in Fig. 5 (b) we find substantial deviations across nearly the entire region considered, although the applicability range of these formulas is more limited. While these parameterizations are fitted to reproduce known experimental data, their extrapolation toward neutron-rich isotopes yields predictions that differ from our microscopic results by more than 40 orders of magnitude.

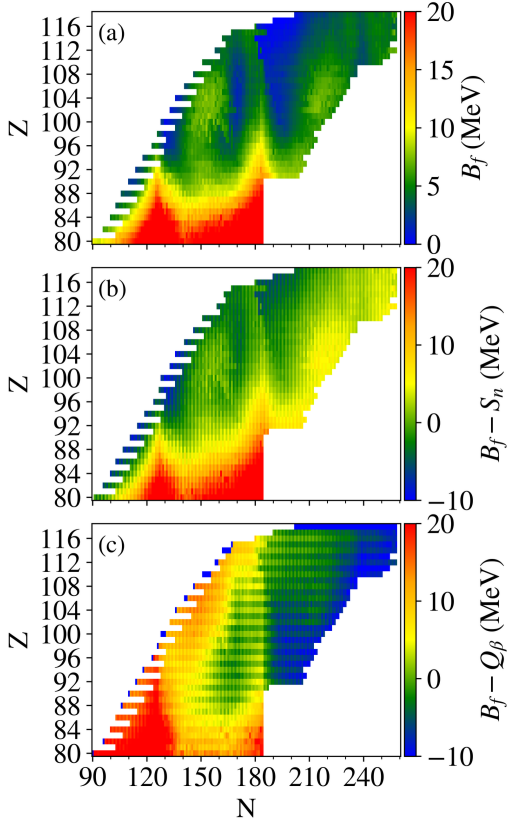

**Fig. 4.** (a) Primary fission barriers, $B_f$ and their difference with (b) the neutron separation energy, $S_n$, and (c) $\beta^-$ $Q$-value in the region $Z = [80, 118]$ obtained with BSkG3.

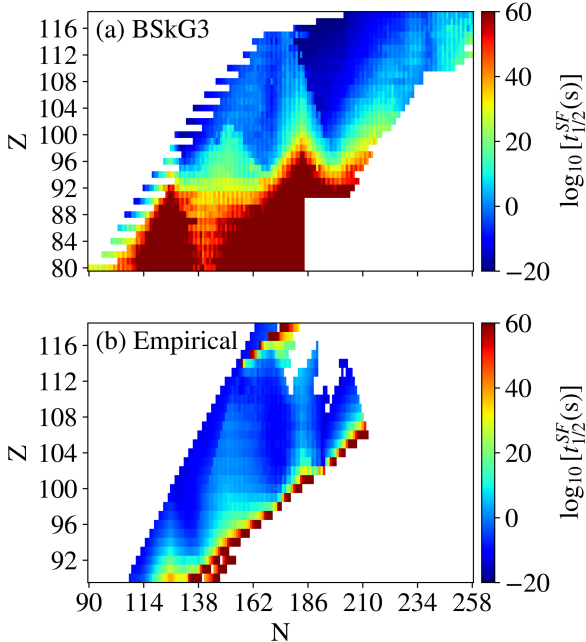

**Fig. 5.** Decimal logarithm of SF half-lives in the region $Z = [80, 118]$ obtained with (a) BSkG3 and (b) empirical formulas from [22] using FRLDM and ETSFI primary fission barriers.

## 4 Conclusions and future work

Taking advantage of the computational efficiency of the coordinate-space representation of the HFB equations and the predictive power of the BSkG3 functional, we have performed systematic fission calculations for more than 3,000 nuclei in the region relevant to the r-process. By computing two-dimensional PESs, we constructed fission paths that maximize the semiclassical barrier penetration probability through minimization of the least-action integral, explicitly including both triaxial and octupole degrees of freedom. Moreover, the explicit treatment of octupole deformation enabled the calculation of fission fragment yields, which showed promisingly good agreement with experimental data. In this sense, BSkG3 represents, to our knowledge, the first microscopic model applied not only to describe GS properties but also to predict large-scale fission properties, including fission paths and fission fragments distributions.

However, this work remains incomplete. While all PESs and fission paths have been generated, a complete set of nuclear inputs for r-process simulations still requires the determination of full fission fragment distributions, including prompt neutron emission effects. In addition, neutron-induced and $\beta$-delayed fission rates must be derived from the computed fission paths, level densities implemented in the TALYS code [25] and QRPA $\beta$-feeding functions. These inputs will ultimately be incorporated into nucleosynthesis calculations in astrophysical conditions such as neutron star mergers.

In previous studies, we have shown that microscopic EDF approaches can achieve an accuracy comparable to (or even better than) highly parametrized phenomenological models. Here, we further demonstrate that such microscopic frameworks can also be applied at scale, providing a consistent and predictive description of nuclear properties relevant for astrophysical applications.

This work was supported by the Fonds de la Recherche Scientifique (F.R.S.-FNRS) and the Fonds Wetenschappelijk Onderzoek – Vlaanderen (FWO) under the EOS Projects No. O000422 and O022818F, as well as by the F.R.S.-FNRS under the MIS Project No. 40028446. This research benefited from computational resources made available on the Tier-1 supercomputer *Lucia* of the Fédération Wallonie-Bruxelles, infrastructure funded by the Walloon Region under Grant Agreement No. 1117545. Additional computational resources were provided by the Consortium des Équipements de Calcul Intensif (CÉCI), funded by F.R.S.-FNRS under Grant No. 2.5020.11 and by the Walloon Region. We acknowledge the EuroHPC Joint Undertaking for awarding access to the *MareNostrum5* supercomputer at the Barcelona Supercomputing Center (BSC), Spain, under projects EHPC-DEV-2025D04-097 and EHPC-DEV-2026D01-043. We further acknowledge EPICURE, a EuroHPC Joint Undertaking initiative, for supporting our work on MareNostrum5 in the context of these development projects.

## References

1. S. Goriely. The fundamental role of fission during r-process nucleosynthesis in neutron star mergers. Eur. Phys. J. A 51, 22 (2015). https://doi.org/10.1140/epja/i2015-15022-3
2. S. A. Giuliani, G. Martínez-Pinedo, L. M. Robledo, Fission properties of superheavy nuclei for r-process calculations. Phys. Rev. C 97, 034323 (2018). https://doi.org/10.1103/PhysRevC.97.034323
3. P. Möller, A. J. Sierk, T. Ichikawa, H. Sagawa, Nuclear ground-state masses and deformations: FRDM(2012). Phys. Rev. C 91, 024310 (2015). https://doi.org/10.1016/j.adt.2015.10.002
4. Rodríguez-Guzmán, R., Humadi, Y.M. & Robledo, L.M. Microscopic description of fission in superheavy nuclei with the parametrization D1M of the Gogny energy density functional. Eur. Phys. J. A 56, 43 (2020). https://doi.org/10.1140/epja/s10050-020-00051-w
5. S. Goriely, G. Martínez-Pinedo, The production of transuranium elements by the r-process nucleosynthesis. Nucl. Phys. A 944, 158–176 (2015). https://doi.org/10.1016/j.nuclphysa.2015.07.020
6. G. Grams, W. Ryssens, G. Scamps, S. Goriely, N. Chamel, Skyrme-Hartree-Fock-Bogoliubov mass models on a 3D mesh: III. From atomic nuclei to neutron stars. Eur. Phys. J. A 59, 270 (2023). https://doi.org/10.1140/epja/s10050-023-01158-6
7. A. Sánchez-Fernández, S. Bara, W. Ryssens, S. Goriely, Accurate spontaneous fission half-lives from a microscopic large-scale nuclear structure model. Phys. Lett. B 874, 140287 (2026). https://doi.org/10.1016/j.physletb.2026.140287
8. A. Koning, S. Hilaire, S. Goriely, TALYS: modeling of nuclear reactions. Eur. Phys. J. A 59, 131 (2023). https://doi.org/10.1140/epja/s10050-023-01038-3
9. W. Ryssens, G. Scamps, S. Goriely, M. Bender, Skyrme–Hartree–Fock–Bogoliubov mass models on a 3D mesh: IIb. Fission properties of BSkG2. Eur. Phys. J. A 59, 96 (2023). https://doi.org/10.1140/epja/s10050-023-01002-x
10. A. Zdeb, M. Warda, L. M. Robledo, Description of the multidimensional potential-energy surface in fission of $^{252}$Cf and $^{258}$No. Phys. Rev. C 104, 014610 (2021). https://doi.org/10.1103/PhysRevC.104.014610
11. M.-H. Zhou, Z.-Y. Li, S.-Y. Chen, Y.-J. Chen, Z.-P. Li, Three-dimensional potential energy surface for fission of $^{236}$U within covariant density functional theory. Chin. Phys. C 47, 064106 (2023). https://doi.org/10.1088/1674-1137/acc4ac
12. W. Ryssens, Symmetry breaking in nuclear mean-field models. PhD Thesis, Université Libre de Bruxelles (2016).
13. E. Flynn, D. Lay, S. Agbemava, P. Giuliani, K. Godbey, W. Nazarewicz, J. Sadhukhan, Nudged elastic band approach to nuclear fission pathways. Phys. Rev. C 105, 054302 (2022). https://doi.org/10.1103/PhysRevC.105.054302
14. A. Sánchez-Fernández et al., in preparation (2026).
15. R. Capote, M. Herman, P. Obložinský, et al, RIPL — Reference Input Parameter Library for Calculation of Nuclear Reactions and Nuclear Data Evaluations. Nucl. Data Sheets 110, 3107–3214 (2009). https://doi.org/10.1016/j.nds.2009.10.004
16. F. G. Kondev, M. Wang, W. J. Huang, S. Naimi, G. Audi, The NUBASE2020 evaluation of nuclear physics properties. Chin. Phys. C 45, 030001 (2021). https://doi.org/10.1088/1674-1137/abddae
17. S. Goriely, M. Samyn, J. M. Pearson, Further explorations of Skyrme-Hartree-Fock-Bogoliubov mass formulas. VII. Simultaneous fits to masses and fission barriers. Phys. Rev. C 75, 064312 (2007).
18. M. Baldo, L. M. Robledo, P. Schuck, X. Viñas, New Kohn-Sham density functional based on microscopic nuclear and neutron matter equations of state. Phys. Rev. C 87, 064305 (2013). https://doi.org/10.1103/PhysRevC.87.064305
19. D. Regnier, N. Dubray, M. Verrière, N. Schunck, FELIX-2.0: New version of the finite element solver for the time dependent generator coordinate method with the Gaussian overlap approximation. Comput. Phys. Commun. 221, 117–133 (2017). https://doi.org/10.1016/j.cpc.2017.12.007
20. D. A. Brown, M. B. Chadwick, R. Capote, et al., ENDF/B-VIII.0: The 8th Major Release of the Nuclear Reaction Data Library with CIELO-project Cross Sections, New Standards and Thermal Scattering Data. Nucl. Data Sheets 148, 1–142 (2018). https://doi.org/10.1016/j.nds.2018.02.001
21. O. Iwamoto, N. Iwamoto, S. Kunieda, F. Minato, S. Nakayama, Y. Abe, et al., "Japanese evaluated nuclear data library version 5: JENDL-5", J. Nucl. Sci. Technol., 60(1), 1-60 (2023). http://doi.org/10.1080/00223131.2022.2141903
22. S. Wanajo, Physical Conditions for the r-process. I. Radioactive Energy Sources of Kilonovae. Astrophys. J. 868, 65 (2018). https://doi.org/10.3847/1538-4357/aae0f2
23. J. Khuyagbaatar, Spontaneous fission half-lives of the heaviest nuclei: Semi-empirical predictions. Nucl. Phys. A 1002, 121958 (2020). https://doi.org/10.1016/j.nuclphysa.2020.121958
24. A. Mamdouh, J. M. Pearson, M. Rayet, and F. Tondeur, Fission barriers of neutron-rich and superheavy nuclei calculated with the ETFSI method, Nucl. Phys. A 679, 337–358 (2001). https://doi.org/10.1016/S0375-9474(00)00358-4
25. S. Goriely, W. Ryssens, S. Hilaire, and A. J. Koning, Improved microscopic nuclear level densities within the triaxial Hartree–Fock–Bogoliubov plus combinatorial method, Phys. Rev. C 113, 014320 (2026). https://doi.org/10.1103/r6pn-xjnv